# Relaxed Belief Propagation for MIMO Detection

Feichi Long, *Student Member, IEEE* and Tiejun Lv, *Member, IEEE*
Key Lab of Universal Wireless Communications, Ministry of Education
Beijing University of Posts and Telecommunications, Beijing, China 100876
{longfish, lvtiejun}@bupt.edu.cn

*Abstract*—In this paper, relaxed belief propagation (RBP) based detectors are proposed for multiple-input multiple-out (MIMO) system. The factor graph is leveraged to represent the MIMO channels, and based on which our algorithms are developed. Unlike the existing complicated standard belief propagation (SBP) detector that considers all the edges of the factor graph when updating messages, the proposed RBP focuses on partial edges, which largely reduces computational complexity. In particular, relax degree is introduced in to determine how many edges to be selected, whereby RBP is a generalized edge selection based BP method and SBP is a special case of RBP having the smallest relax degree. Moreover, we propose a novel Gaussian approximation with feedback information mechanism to enable the proposed RBP detector. In order to further improve the detection performance, we also propose to cascade a minimum mean square error (MMSE) detector before the RBP detector, from which pseudo priori information is judiciously exploited. Convergence and complexity analyses, along with the numerical simulation results, verify that the proposed RBP outperform other BP methods having the similar complexity, and the MMSE cascaded RBP even outperform SBP at the largest relax degree in large MIMO systems.

## I. INTRODUCTION

Although belief propagation (BP) is rooted in artificial intelligence, it has attracted vast attentions among communication theorists since the argument that decoding of the celebrated turbo code as an instance of BP [1]. In the past decade, graph models and BP have been embedded in many contexts of communication literature, such as coding theory [2], equalization, multi-user detection [3] and signal processing [4]. One of the most concerned issues is that for graphs having many circles, BP does not converge, thereby poor performance and high complexity become the main drawbacks [5].

We consider the detection problem in multi-input multi-out (MIMO) systems. Various detection algorithms have long been proposed like zero forcing (ZF), minimum mean square error (MMSE) and serial interference cancellation (SIC). In terms of BP's compatibility with other problems such as decoding and channel estimation, BP has been applied for MIMO detection recently [6]-[9]. The graph model of MIMO channels is a complete graph where each variable is connected to all the other variables, which means the graph has the most circles among the graphs having the equivalent variables. Standard BP in these complete graphs achieves poor performance [6], [7], or possesses high complexity [8]. Thereby designing message

This work is financially supported by the National Natural Science Foundation of China (NSFC) under Grant No. 60972075.

and message update rule of BP become the main issues for achieving good performance with low complexity.

The main contribution of our work is that we develop the present BP based algorithms, bring in edge selection policy, some approximation ideas and obtain some methods which outperform other BP based methods. [6] proposed an algorithm based on markov random field (MRF) model, but with very poor performance (see Fig.3). [8] proposed the factor graph model of MIMO channels and the corresponding standard BP algorithm, although it can achieve near optimal performance, the complexity of it is even higher than maximum likelihood (ML) detector. Some low complexity BP algorithms based on edge selection are proposed in [8] as well, while the edge selection policy and approximation method need to be improved. Some other low complexity algorithms are also proposed recently in [7], [9], we take the cue from these algorithms, combined with our original ideas to design the message on factor graph, and modify some key points. Moreover, the message update rule and edge selection policy of [8] is redesigned by us. In particular, relax degree is introduced in to determine how many edges to be selected, whereby the proposed RBP is an adaptive method for tradeoff between performance and complexity.

Numerical simulation results under different antennas, different relax degree and different iterations are analyzed and compared. Meanwhile, convergence are also analyzed with the help of average mutual information (AMI).

The rest of this paper is organized as follows: Section II describes a system model. Section III presents the factor graph representation of MIMO channels and a brief review of standard BP detector. Section IV introduces the proposed relaxed BP algorithms and their performance simulation results. Convergence and complexity analyses are presented in Section V. Finally, conclusions are provided in Section VI.

## II. SYSTEM MODEL

Supposing there are $N_t$ transmit antennas and $N_r$ receive antennas, data sequence is separately modulated and sent through each transmit antenna. Frequency-flat fading channel is considered here and the channel is formulated as

$$\mathbf{y} = \mathbf{H} \cdot \mathbf{s} + \mathbf{n}, \tag{1}$$

where $\mathbf{y}_{N_r \times M} = (y_1, y_2, \cdots, y_{N_r})^T$ is the received vector, $\mathbf{H}_{N_r \times N_t} = \left[\mathbf{h}_1^T, \mathbf{h}_2^T, \cdots \mathbf{h}_{N_r}^T\right]^T$ is the matrix representation of the channel with the elements $h_{j,k}$ denoting the channel gain from transmit antenna $k$ to receive antenna

$j$ and has the distribution $h_{j,k} \sim \mathcal{CN}(0,1)$, $\mathbf{n}_{N_r \times M} = (n_1, n_2, \cdots, n_{N_r})^T$ is additive Gaussian noise vector whose elements satisfy $\mathcal{CN}(0, \sigma^2)$, $\mathbf{s}_{N_t \times M} = (s_1, s_2, \cdots, s_{N_t})^T$ is the transmitted symbol matrix with each symbol independently selected from a set of constellation $\mathcal{A}^M$. The transmitted bit vector is thus $\mathbf{x} = (x_1, x_2, \cdots, x_{MN_t})^T$, where $s_k = (x_{kM-M+1}, x_{kM-M+2}, \cdots, x_{kM})^T$, $k = 1, 2, \cdots, N_t$ and $x_i \in \{+1, -1\}$, $i = 1, 2, \cdots, MN_t$. The $j$th receive signal is then expressed as

$$y_j = \mathbf{h}_j \cdot \mathbf{s} + n_j = \sum_{i=1}^{MN_t} h_{j,k(i)} \cdot x_i + n_j, \quad (2)$$

where $k(i) = 1 + \lfloor (i - 0.5)/M \rfloor$ denotes the subscript of channel gain for bit $x_i$.

We assume that all channel coefficients are independent from each other and are known only at the receiver side.

## III. GRAPH MODELS AND STANDARD BP

### A. Factor Graph for MIMO channels

Factor graphs can describe all the relationships between variables easily, one for MIMO channels is represented in Fig.1, bit nodes $\{x_1, x_2, \cdots, x_{MN_t}\}$ correspond to the transmitted bits, factor nodes $\{f_1, f_2, \cdots, f_{N_r}\}$ represent the relationships between received signal and transmitted bit nodes.

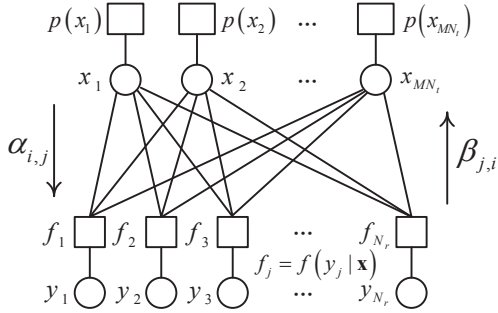

Figure 1. Factor graph model for MIMO channels, where $\alpha$ and $\beta$ are messages passing between the bit nodes and factor nodes.

### B. Standard BP on Factor Graph (SBP)

The essentials of SBP on factor graph lie in two steps at each iteration, first each bit node passes its a priori message to the connected factor nodes, second each factor node updates its a posteriori message according to a priori messages, and then passes it back to each bit node. When bit nodes received the messages from factor nodes, they update their a priori information as well. In the following analyses, log likelihood ratios (LLR) are used for message design, let $j = 1, 2, \cdots, N_r$, $i = 1, 2, \cdots, MN_t$, $\mathbf{x}^{n \backslash t} \triangleq (x_1, \cdots, x_{t-1}, x_{t+1}, \cdots, x_n)$ and $\mathcal{R}(x)$ denotes the real part of $x$.

*1) Message Design:* For each bit nodes, LLR of each bit's a priori probability in $\mathbf{s}$ is used as the message $\alpha$ which passes to the factor nodes. The message passing from $x_i$ to $f_j$ at $l$th iteration is

$$\alpha_{i,j}^{(l)} = \log \frac{p^{(l)}(x_i = 1)}{p^{(l)}(x_i = -1)}. \quad (3)$$

Assume that input bits are independent, then at the $l$th iteration, the message sent from $f_j$ to $x_i$, representing a posteriori probability of signal and denoted by $\beta$, can be calculated as follows:

$$\beta_{j,i}^{(l)} = \log \frac{p(x_i = 1 \mid y_j, \mathbf{h}_j)}{p(x_i = -1 \mid y_j, \mathbf{h}_j)}$$
$$= \log \frac{\sum_{\mathbf{s}:x_i=1} p(y_j \mid \mathbf{s}, \mathbf{h}_j) \cdot p^{(l)}(\mathbf{x}^{MN_t \backslash i})}{\sum_{\mathbf{s}:x_i=-1} p(y_j \mid \mathbf{s}, \mathbf{h}_j) \cdot p^{(l)}(\mathbf{x}^{MN_t \backslash i})}. \quad (4)$$

Using the approximation $\log(e^x + e^y) \approx \max(x, y)$, we get:

$$\beta_{j,i}^{(l)} \approx \max_{\mathbf{s}:x_i=1} \left\{ D_j(\mathbf{s}) + \sum_{t:x_t=1, t \neq i} \alpha_{t,j}^{(l-1)} \right\}$$
$$- \max_{\mathbf{s}:x_i=-1} \left\{ D_j(\mathbf{s}) + \sum_{t:x_t=1, t \neq i} \alpha_{t,j}^{(l-1)} \right\}, \quad (5)$$

where the function $D_j(\mathbf{s})$ is defined as

$$D_j(\mathbf{s}) \triangleq \log(p(y_j \mid \mathbf{s}, \mathbf{h}_j)) \propto -\frac{1}{2\sigma^2} \| y_j - \mathbf{h}_j \mathbf{s} \|^2. \quad (6)$$

*2) Message Update Rule:* (5) actually gives the message update rule of $\beta$, on the other hand, the update rule of $\alpha$ is

$$\alpha_{i,j}^{(l)} = \sum_{t=1, t \neq j}^{N_r} \beta_{t,i}^{(l-1)}, \quad (7)$$

which is the standard BP message update rule.

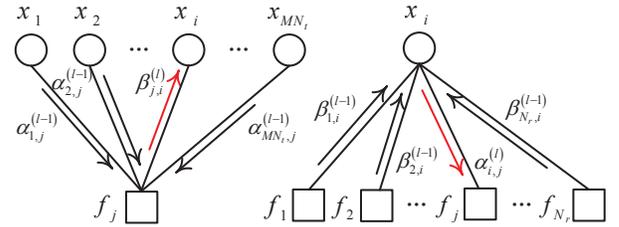

Figure 2. Message update rule of SBP on factor graph at $l$th iteration. The left represents message updating of $\beta_{j,i}^{(l)}$ with all the $\alpha$ obtained at $(l-1)$th iteration, while the right represents message updating of $\alpha_{i,j}^{(l)}$.

Fig.2 describes the message update rule of SBP. As can be seen, all edges participate in message updates, each edge passes extrinsic information obtained in the last iteration from its neighbor edges, and then send updated message to all the connected bit or factor nodes.

*3) Soft Output:* Soft computation after $L$ times iteration takes place as

$$S_{out}(x_i) = \sum_{t=1}^{N_r} \beta_{t,i}^{(L)}, \quad (8)$$

and detection result of bit $x_i$ is $\hat{x}_i = \text{sgn}(S_{out}(x_i))$.

*4) Performance of SBP Detector:* Bit error rate (BER) simulation results of SBP and BP on MRF defined in [6] are shown in Fig.3, from which one can easily draw out that SBP can achieve near optimal performance.

*5) Complexity Drawbacks:* Although SBP achieves near optimal performance, the calculated amounts for per channel use of SBP is even higher than ML (see Tab.I), which seriously prohibits for its use in practical scenarios.

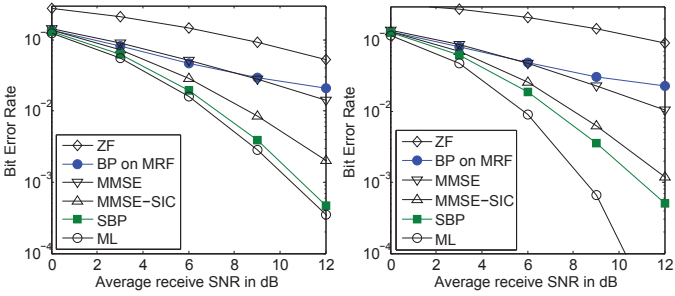

Figure 3. BER performance comparison of BP based detectors in [6] (BP on MRF) and in [8] (SBP), BPSK modulation, $L=5$, $N_t = N_r = 4$ of the left while $N_t = N_r = 8$ of the right.

## IV. RELAXED BP

### A. Edge Selection Policy for Complexity Reduction

From (5) and (7), one can easily draw out that calculation complexity mostly lies in the message updating of $\beta$, while message updating of $\alpha$ contains only $N_r - 1$ additons and contribute little complexity. As represented in Fig.2, when updating $\beta_{j,i}^{(l)}$ from $f_j$ to $x_i$ at $l$th iteration, all the neighbor edges participate in and pass $\alpha_{t,j}^{(l-1)}$ obtained in the last iteration, where $t = 1, 2, \cdots, MN_t$ and $t \neq i$.

We propose that for complexity reduction, only $R_{d_1} M + R_{d_2}(M-1)$ neighbor edges are selected when updating $\beta_{j,i}^{(l)}$, where $R_{d_1}$ is a coefficient correlated with relax degree (defined as $N_t - R_{d_1} - 1$) and $R_{d_1} = 0, 1, \cdots, (N_t - 1)$, $R_{d_2}$ is another coefficient and $R_{d_2} = 0, 1$. The $R_{d_1} M$ neighbor edges correspond to $R_{d_1}$ largest sub-channel gains among all the $N_t$ elements in $\mathbf{h}_j$ except the sub-channel gain $h_{j,k(i)}$, and $R_{d_2}(M-1)$ neighbor edges are the edges with the same sub-channel gain $h_{j,k(i)}$. In the following context, $\Psi_{j,i} \triangleq \left\{ \psi_1^{j,i}, \psi_2^{j,i}, \cdots, \psi_{R_D}^{j,i} \right\}$ denotes the set containing subscripts of the $R_D$ ($R_D = R_{d_1} M + R_{d_2}(M-1)$) neighbor edges, $\mathbf{s}'$ denotes the vector which contains the bits corresponding to the selected edges and RBP($R_{d_1}, R_{d_2}$) denotes the relaxed BP algorithm with coefficient $R_{d_1}$ and $R_{d_2}$.

Under the edge selection policy, (5) becomes

$$\beta_{j,i}^{(l)} \approx \max_{\mathbf{s}': x_i = 1} \left\{ D_j(\mathbf{s}') + \sum_{t \in \Psi_{j,i}, x_t = 1} \alpha_{t,j}^{(l-1)} \right\} - \max_{\mathbf{s}': x_i = -1} \left\{ D_j(\mathbf{s}') + \sum_{t \in \Psi_{j,i}, x_t = 1} \alpha_{t,j}^{(l-1)} \right\}, \quad (9)$$

where $D_j(\mathbf{s}')$ is the primary issue we need to approximate.

Two extreme cases are RBP($N_t - 1, 1$) and RBP(0,0), the former turns out to be SBP and the latter is the most simplified, we will discuss RBP(0,0) alone, and in this case

$$\beta_{j,i}^{(l)} = D_j(\mathbf{s}' \mid \mathbf{s}' : x_i = 1) - D_j(\mathbf{s}' \mid \mathbf{s}' : x_i = -1). \quad (10)$$

The message update rule of $\beta_{j,i}^{(l)}$ in RBP is shown in Fig.4, the dashed line means that there is no message passed on this edge. What's more, the message update rule of $\alpha_{i,j}^{(l)}$ in RBP is the same with SBP as shown in the right part of Fig.2.

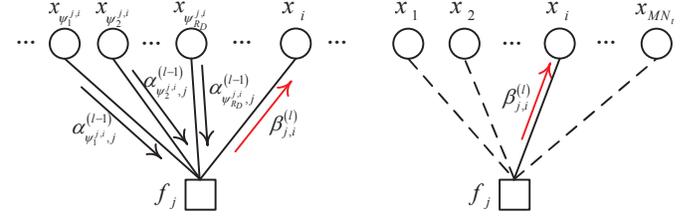

Figure 4. Message update rule of $\beta_{j,i}^{(l)}$ in RBP at $l$th iteration. The left represents message updating in RBP with only $R_D$ of the $\alpha$ obtained at $(l-1)$th iteration, the right represents the case RBP(0,0).

### B. Gaussian Approximation with Feedback Information

Under the edge selection policy, when updating $\beta_{j,i}^{(l)}$ and calculating $D_j(\mathbf{s}')$, the signal model is represented by

$$y_j = h_{j,k(i)} x_i + \sum_{t \in \Psi_{j,i}} h_{j,k(t)} x_t + \sum_{t \neq i, t \notin \Psi_{j,i}} h_{j,k(t)} x_t + n_j$$

$$\triangleq h_{j,k(i)} x_i + \sum_{t \in \Psi_{j,i}} h_{j,k(t)} x_t + z_{j,i}, \quad (11)$$

where $z_{j,i}$ is viewed as 'interference' or noise, as in the literature of multi-user detection. Gaussain approximation is applied here as $z_{j,i} \sim \mathcal{CN}\left(u_{z_{j,i}}, \sigma_{z_{j,i}}^2\right)$.

The key points in this sorts of algorithms are the modeling of mean and variance. We argue that a priori messages obtained in the last iteration can be adopted in the next iteration as feedback information. Therefore $u_{z_{j,i}}$ can be formulized as

$$u_{z_{j,i}}^{(l)} = \sum_{t \neq i, t \notin \Psi_{j,i}} h_{j,k(t)} \cdot \mathbb{E}\left(\hat{x}_t^{(l)}\right)$$

$$= \sum_{t \neq i, t \notin \Psi_{j,i}} h_{j,k(t)} \cdot \left(2p^{(l)}(x_t = 1) - 1\right) \quad (12)$$

at $l$th iteration, where $p^{(l)}(x_t)$ can be calculated by the message $\alpha_{t,j}^{(l)}$ (see (3)). And for the variance, we formulize as

$$\sigma_{z_{j,i}}^2 = \sum_{t \neq i, t \notin \Psi_{j,i}} |h_{j,k(t)}|^2 \cdot \text{var}(x_t) + \sigma^2$$

$$= \sum_{t \neq i, t \notin \Psi_{j,i}} |h_{j,k(t)}|^2 + \sigma^2, \quad (13)$$

here the results of last iteration are not used for $\text{var}(x_t)$ in view of the fact that the approximation of $x_t$'s variance is more sensitive to the output and only variance of the priori distribution before the first iteration is enough. The modeling of mean and variance is different from [9], where they update both.

After modeling of (12) and (13), the function $D_j(\mathbf{s}')$ in (9) is thus

$$D_j(\mathbf{s}') = -\frac{1}{2\sigma_{z_{j,i}}^2} \| y_j - h_{j,k(i)} x_i - \sum_{t \in \Psi_{j,i}} h_{j,k(t)} x_t - u_{z_{j,i}}^{(l-1)} \|^2. \quad (14)$$

Thereafter $\beta_{j,i}^{(l)}$ can be calculated out from (14) and (9). Especially, for RBP(0,0) there is

$$\beta_{j,i}^{(l)} = \frac{2}{\sigma_{z_{j,i}}^2}\mathcal{R}\left(h_{j,k(i)}^*\left(y_j - u_{z_{j,i}}^{(l-1)}\right)\right). \quad (15)$$

As for message $\alpha$, the updating rule is the same as (7).

*C. MMSE Detector Cascaded RBP*

In [7] the author proposed a algorithm that use pseudo priori information obtained from linear detector, while this algorithm is essentially an improved BP based detector on MRF. Since BP on MRF achieves poor performance, we take the cue from [7]'s algorithm and apply it into our factor graph model.

The main idea is to initialize all the message $\alpha$ by just one MMSE detector, that is, a priori message before first iteration $\alpha_{i,j}^{(0)}$. MMSE detector is a linear filter defined as

$$\hat{\mathbf{s}}_{MMSE} = \left(\mathbf{H}^*\mathbf{H} + \sigma^2 \cdot \mathbf{I}\right)^{-1} \cdot \mathbf{H}^* \cdot \mathbf{y}, \quad (16)$$

which has comparatively very low complexity (here $\mathbf{I}$ is a $N_t \times N_t$ unit matrix). Thereby the priori distribution of $\mathbf{x}$ (initial value when $L=0$) can be approximated as

$$p^{(0)}(x_i) \propto \exp\left(-\frac{\|x_i - \hat{\mathbf{s}}_{MMSE_{k(i)}}\|^2}{2\sigma_{MMSE_{k(i)}}^2}\right), \quad (17)$$

where $\hat{\mathbf{s}}_{MMSE_{k(i)}}$ is the $k(i)$th element of $\hat{\mathbf{s}}_{MMSE}$ defined in (16), and $\sigma_{MMSE_{k(i)}}^2$ is the element of the following nosie covariance matrix in $k(i)$th row and $k(i)$th column:

$$\mathbf{K}_{z_{MMSE}} = \left(\mathbf{H}^*\mathbf{H} + \sigma^2 \cdot \mathbf{I}\right)^{-1}. \quad (18)$$

Above this, a priori message is

$$\alpha_{i,j}^{(0)} = \frac{2\mathcal{R}\left(\hat{\mathbf{s}}_{MMSE_{k(i)}}\right)}{\sigma_{MMSE_{k(i)}}^2}. \quad (19)$$

After each a priori message of $\alpha_{i,j}^{(0)}$ is obtained from MMSE filter, the messages pass to the factor nodes and iterate $L$ times in the same way of RBP, and this is called MMSE-RBP.

*D. Performance of RBP and MMSE-RBP*

We provide simulation results for the proposed algorithms, which are shown in Fig.5 and Fig.6. BPSK modulation is adopted here, thus $R_{d_2}$ is insignificant and not considered. [8] also proposed an edge based BP algorithm, denoted by EB, we compare it with RBP at different $R_{d_1}$. As can be seen, the proposed RBP achieves much better performance than EB in [8], while the complexity is no higher, as will be shown latter. What's more, as SBP is just RBP($N_t-1,1$), we can point out that the BER of RBP will approach SBP's asymptotically as $R_{d_1}$ increase, which is partly shown in the figures.

As for MMSE-RBP detector, its performance is nearly the same as MMSE-SIC when SNR belows 12dB in $4 \times 4$ systems, while it is so amazing that MMSE-RBP even outperforms SBP both when $R_{d_1}=0$ and when $R_{d_1}=1$ in $8 \times 8$ systems. Despite there are error floors due to approximation and edge selection, the performance of MMSE-RBP is much satisfactory in view of the fact that its complexity is much lower than SBP.

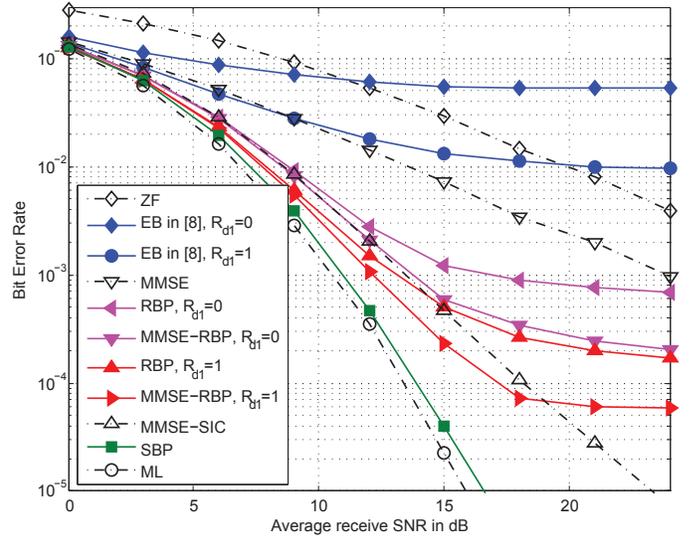

Figure 5. BER comparison of different BP based detectors, BPSK modulation, $N_t = N_r = 4$, $L = 7$.

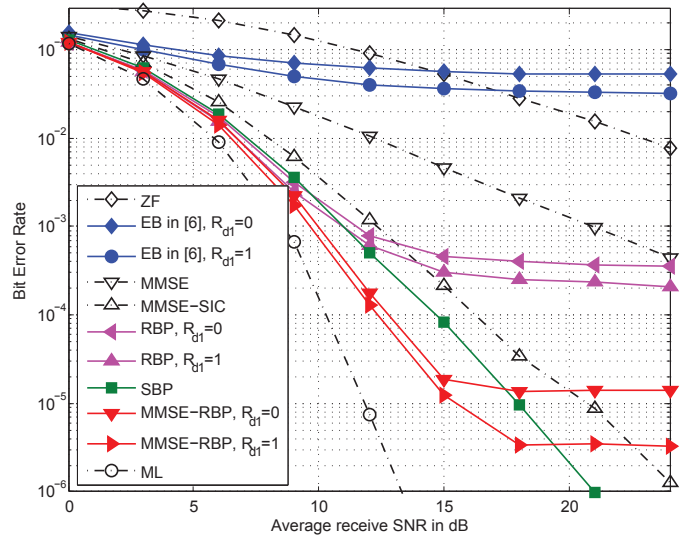

Figure 6. BER comparison of different BP based detectors, BPSK modulation, $N_t = N_r = 8$, $L = 5$.

When numble of circles becomes large, BP on loopy graph converges very slowly and the performance can not be guaranteed, which is still an open problem. Intuitionistic explanation for the excellent performance of MMSE-RBP at low SNR is that, it cut off all the 'redundant' edges and forms a dynamic graph which has much less circles, while SBP passes mesage on all the edges and in both directions.

## V. COMPLEXITY AND CONVERGENCE ANALYSES

Table.I gives comparison of computational complexity of the detectors mentioned above, where $\mathcal{C}$ (RBP) denotes the complexity of RBP in this item. One can see that SBP has even higher complexity than ML, while RBP and MMSE-RBP have much lower complexity if $R_{d_1}$ is small.

For convergence, average mutual information (AMI) is one of the most accurate measures for tracking the convergence behavior of iterative systems, the higher AMI is, the more reliable the detector is. Readers can refer to [8] for details about AMI. Simulation results of AMI are shown in Fig.7, from which we can see that due to disconnectivity of the graph, the AMI of the proposed RBP detectors are smaller than that of SBP. Furthermore, MMSE-RBP has larger AMI compared with RBP because of the additional a priori information.

Table I
COMPLEXITY OF DIFFERENT ALGORITHMS PER CHANNEL USE

| Algorithm | multiplications | additions | comparisons |
|---|---|---|---|
| ML | $2^{MN_t} N_r N_t + M N_t$ | $2^{MN_t} N_r N_t + 2^{MN_t} M N_t$ | 0 |
| SBP | $2^{MN_t} N_r N_t$ | $\left(2^{MN_t} + \left(2^{MN_t-1} + 3\right) ML\right) N_r N_t$ | $\left(2^{MN_t} - 2\right) M N_t N_r L$ |
| RBP | $2^{R_D+1} (R_{d_1}+1) N_r$ $+ 2^M (N_t - R_{d_1} - 1) N_r$ | $2^{R_D+1} (R_{d_1}+1) N_r + 2^M (N_t - R_{d_1} - 1) N_r$ $+ \left(2^{R_D}(R_{d_1}+1) + 2^{M-1} + 3N_t\right) M L N_r$ | $\left(2^{R_D+1} - 2\right) M L N_t N_r$ |
| MMSE-RBP | $\mathcal{C}(\text{RBP}) + O(N_t^3)$ | $\mathcal{C}(\text{RBP}) + O(N_t^3)$ | $\mathcal{C}(RBP) + N_t(N_t-1)/2$ |
| RBP(0,0) | $2N_r + 2^M (N_t - 1) N_r$ | $2N_r + 2^M (N_t - 1) N_r + \left(2^{M-1} + 3N_t + 2\right) M L N_t N_r$ | 0 |
| EB in [8] | $2^{MR_{d_1}+M} (R_{d_1}+1) N_r$ $+ 2^M (N_t - R_{d_1} - 1) N_r$ | $2^{MR_{d_1}+M} (R_{d_1}+1) N_r + 2^M (N_t - R_{d_1} - 1) N_r$ $+ \left(2^{MR_{d_1}+M-1}(R_{d_1}+1) + 2^{M-1} + 3N_t\right) M L N_r$ | $\left(2^{MR_{d_1}+M} - 2\right) M L N_t N_r$ |

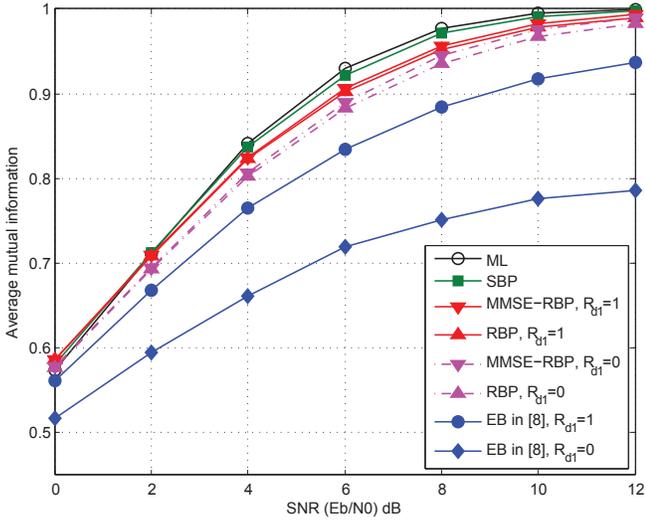

Figure 7. AMI comparison of different BP based detectors, BPSK modulation, $N_t = N_r = 4$, $L = 5$.

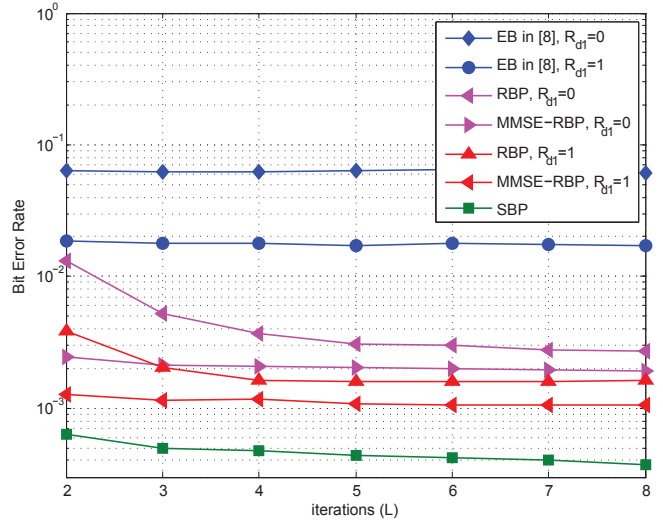

Figure 8. Convergence comparison of different BP based detectors, BPSK modulation, $N_t = N_r = 4$, SNR = 12dB.

Besides the semi-analytical method of AMI, we give the convergence comparison of different BP based detectors with different iteration number, which is shown in Fig.8. The convergence speed is different but $L = 5$ is enough for almost all algorithms in practical implementation.

## VI. CONCLUSIONS

Form the above analysis, the proposed RBP detectors can achieve high performance with low complexity at large relax degree, and achieve near optimal performance asymptotically when the relax degree decrease. The special case RBP(0,0) possesses smallest complexity and achieves satisfactory performance if MMSE filter is cascaded, which is a iterative parallel interference cancellation method essentially, we will analyze it alone theoretically in another paper.